\documentclass{article}
\usepackage{epsfig}
\begin{document}
\begin{center}
{\Large\bf The Gordon-Haus effect for modified NLS solitons}

\bigskip
E.V. Doktorov and I.S. Kuten

\bigskip
   B.I. Stepanov Institute of Physics \\
  68 F. Skaryna Ave., 220072 Minsk, Belarus
\end{center}
\begin{abstract}
Random jitter in the soliton arrival time (the Gordon-Haus
effect) is analyzed for solitons being solutions of the
integrable modified nonlinear Schr\"odinger equation. It is shown
that the mean square fluctuation of the soliton position depends
on the soliton parameters which can be properly adjusted to
suppress the Gordon-Haus jitter.
\end{abstract}

\section{Introduction}
Long-distance soliton-based fibre transmission meets with
difficulties imposed by random jitter in the soliton arrival time
caused by the spontaneous amplifier noise (the so-called
Gordon-Haus effect~\cite{1}).The amplifier noise incorporated by
the soliton produces a random soliton frequency shift which leads
to a timing shift because of the group velocity dispersion. The
Gordon-Haus effect limits error-free propagation of the soliton.
There is a few ways to partially overcome the Gordon-Haus limit,
including the use of linear filtering~\cite{2,3}, the dispersion
compensation means~\cite{4,5,6}, the use of a sequence of two
different media to reduce a path-averaged dispersion~\cite{7,8}.

From the viewpoint of the inverse scattering transform, the
theory of the Gordon-Haus effect is a direct consequence of the
adiabatic soliton perturbation theory for the nonlinear
Schr\"odinger (NLS) equation, with perturbation being the
amplifier noise~\cite{9}. The NLS equation serves as the
integrable model describing the picosecond soliton dynamics in
fibres. On the other hand, when dealing with ultrashort optical
pulses with duration $\leq$ 100 fs, the NLS equation should be
modified to adopt more subtle effects, such as the nonlinearity
dispersion, the Raman self-frequency shift and the third-order
dispersion~\cite{10}. It is remarkable that the account for the
nonlinearity dispersion does not break the integrability of the
equation. In other words, the modified NLS (MNLS) equation

\begin{equation}
\label{1}
iu_{z}+\frac{1}{2}u_{tt}+|u|^{2}u+i\alpha(|u|^{2}u)_{t}=0.
\end{equation}
is still integrable by means of the inverse scattering
transform~\cite{11}, though the linear spectral problem associated
with the MNLS equation differs from that for the NLS equation
(with $\alpha=0$). Here $u$ is the normalized slowly varying
amplitude of the electric field envelope, $z$ and $t$ are the
normalized propagation distance and time in the frame comoving
with the group velocity, the real parameter $\alpha$ governs the
effect of the nonlinearity dispersion. We consider the MNLS
equation~(\ref{1}) as the integrable model  for ultrashort
optical pulses, i.e., playing the same role as the NLS equation
does for picosecond solitons. Thereby, we change the status of the
nonlinearity dispersion term from being a perturbation in the NLS
equation to the essential ingredient of the MNLS equation. It is
important that such a change is in no way an issue of our
convenience. It was shown by Ohkuma {\it et al.}~\cite{12} that
numerical simulation of soliton propagation revealed a number of
features which cannot be accounted for by treating this term as a
perturbation term in the NLS equation. In the recent paper
\cite{13}, the adiabatic perturbation theory for MNLS solitons
was elaborated. So, a question to analyze the Gordon-Haus effect
for the MNLS solitons arises naturally.

In this communication, we derive analytically the mean-square
displacement fluctuation for the MNLS soliton propagating in a
fibre. It follows from our results that the Gordon-Haus effect
for the MNLS solitons, as distinct from the NLS solitons, can be
suppressed by properly adjusting the soliton/fibre parameters,
without making use of external means. In the end of the paper we
briefly discuss the suitability of this result to actual
femtosecond solitons.
\section{Formulation of the problem}
We consider the perturbed MNLS equation

\begin{equation}
\label{2}
iu_{z}+\frac{1}{2}u_{tt}+|u|^{2}u+i\alpha(|u|^{2}u)_{t}=s(z,t) ,
\end{equation}
where $s(z,t)$ stands for spontaneous amplifier noise. In order
to digress the details of minor importance, we consider
distributed gain that exactly compensates for the fibre loss.
Further, we consider the amplifier bandwidth to be much larger
than the soliton bandwidth. Because erbium-doped amplifiers have
a gain bandwidth of the order of 40 nm~\cite{9}, this assumption
works well for the ultrashort pulses. We consider the noise
$s(z,t)$ as the delta-correlated function both in time and space,
\begin{equation}
\label{3}
<\bar{s}(z,t)s(z^{\prime},t^{\prime})>=A\delta(t-t^{\prime})\delta(z-z^{\prime})
.
\end{equation}
The estimation of the coefficient $A$ can be found in~\cite{9}.
Finally, we treat the noise $s(z,t)$ as being small to justify the
perturbative approach.

The unperturbed soliton of the MNLS equation~(\ref{1}) was derived
as early as in 1983 by Gerdjikov and Ivanov~\cite{14}. We will use
more simple and transparent expression~\cite{13} for the soliton
solution of~(\ref{1}) :
\begin{equation}
u_{s}(z,t)=\frac{i}{w}\frac{ke^{-x}+\bar{k}e^{x}}{(ke^{x}+\bar{k}e^{-x}
)^{2}}e^{i\psi}. \label{4}
\end{equation}
Here $x$ and $\psi$ are linearly expressed through coordinates $z$
and $t$ :
\begin{equation}
x   =-\frac{t}{w}+q(z), \qquad \psi  =vt+\phi(z) , \qquad q(z)
=a+\frac{v}{w}z, \qquad \phi(z)
=\varphi-\frac{1}{2}(v^{2}-\frac{1}{w^{2}})z \noindent. \label{5}
\end{equation}
Parameters $a$ and $\varphi$ determine initial position and
initial phase of the soliton, the complex parameter $\
k=\xi-i\eta$, $\xi,\eta>0$ determines velocity (more exactly,
shift of the reciprocal soliton velocity) $v$ and
width $w$ of the soliton,%
\begin{equation}
v(z)=\frac{1}{\alpha}-\frac{2}{\alpha}\left(k^2+\bar k^2\right),
\qquad w(z)=-\frac{i\alpha}{2(k^2-\bar k^2)}, \label{6}
\end{equation}
where the $z$-dependence of the soliton velocity and width arises
from the perturbation-induced $z$-dependence of the parameter $k$
(see below~(\ref{8})). Functions $q(z)$ and $\phi(z)$ are the
direct analogs of the Gordon's position and phase parameters for
the NLS soliton~\cite{15}.

The soliton~(\ref{4}) has a number of peculiarities which
distinguish it from the NLS soliton~\cite{13}. For example, the
important invariant of
eq.~(\ref{1}), namely, the  optical energy%
\[
E=\int\limits_{-\infty}^{\infty}dt|u|^{2}=\frac{4}{\alpha}\gamma
\qquad \gamma={\rm Arg}(\bar{k}), \qquad 0<\gamma<\pi/2,
\]
has the upper limit $2\pi/\alpha$. It should be stressed that
$\alpha$ enters the denominator of the soliton solution~(\ref{4})
(because of $w$) , so we account non-perturbatively  for the
nonlinearity dispersion. In other words, the parameter $\alpha$
does not to be small in general. Nevertheless, a non-trivial
limiting procedure exists~\cite{13} permitting to restore
from~(\ref{4}) the NLS soliton as $\alpha\to 0$. Indeed, let us
take the limit
\begin{equation}
k=\frac{1}{2}-\frac{\alpha}{2}k_{\rm NLS}+{\cal O}(\alpha^2),
\qquad k_{\rm NLS}=\xi_{\rm NLS}+i\eta_{\rm NLS}. \label{7}
\end{equation}
Then it is easy to see that the soliton velocity and
width~(\ref{6}) transform in the limit~(\ref{7}) as $v\to
2\xi_{\rm NLS}$, $w\to 1/2\eta_{\rm NLS}$, i. e., to the NLS
parameters, while the MNLS soliton ~(\ref{4}) produces exactly
the NLS soliton $u_{\rm NLS}=2i\eta_{\rm NLS}\exp(i\psi_{\rm
NLS}){\rm sech}x_{\rm NLS}$ with $x_{\rm NLS}=a-2\eta_{\rm
NLS}(\tau-2\xi_{\rm NLS}z)$, $\psi_{\rm NLS}=\phi+2\xi_{\rm
NLS}\tau-2(\xi_{\rm NLS}^2-\eta_{\rm NLS}^2)z$.

As it was shown in~\cite{13}, a perturbation-induced z-evolution
of the parameter
$k$ is given by the simple formula:%
\begin{equation}
\frac{dk}{dz}=\frac{i}{2}\alpha k^{2}\int\limits_{-\infty}^{\infty}dx\frac{e^{x}%
}{(ke^{-x}+\bar{k}e^{x})^{2}}[s(x)e^{-i\psi}+\bar{s}(-x)e^{i\psi}]
.\label{8}
\end{equation}
which is transformed in accordance with~(\ref{6}) into the corresponding relations
for velocity and width:%
\begin{equation}
\label{9}
\frac{dv}{dz}   =-2i\int\limits_{-\infty}^{\infty}dx\frac{k^{3}e^{x}%
-\bar{k}^{3}e^{-x}}{(ke^{-x}+\bar{k}e^{x})^{2}}[s(x)e^{-i\psi}+\bar{s}
(-x)e^{i\psi}]\equiv S_{v}(z),
\end{equation}
\[
 \frac{d}{dz}\frac{1}{w}
=-2\int\limits_{-\infty}^{\infty}dx\frac
{k^{3}e^{x}+\bar{k}^{3}e^{-x}}{(ke^{-x}+\bar{k}e^{x})^{2}}[s(x)e^{-i\psi
}+\bar{s}(-x)e^{i\psi}]\equiv S_{w}(z) .
\]
Here $S_{v}(z)$ and $S_{w}(z)$ stand for the noise sources driving
velocity and width. Because noise $s(z,t)$ is given in terms of
the correlation function~(\ref{3}), the responses are also
expressed in the same form.

We are mostly interested in the mean-square fluctuation $\langle
q(L)q(L) \rangle$ ~($L$ being a fibre length) of the soliton
displacement $q(z)$~(\ref{5}) which is given in the presence of
perturbation by
\[
q(z)=a+\int\limits^{z}dz^{\prime}v(z^{\prime})w^{-1}(z^{\prime}) .
\]
with $v$ and $\ w$ determined from~(\ref{9}). The jitter of the
soliton displacement in the comoving frame is calculated from
\begin{equation}
 \frac{dq}{dz}=\frac{da}{dz}+\frac{v}{w}, \label{10}
\end{equation}
where z-evolution of $a$ is given by \cite{13}%
\begin{equation}
\label{11}
\frac{da}{dz}
=\frac{da_{+}}{dz}+\frac{da_{-}}{dz}\equiv S_{a}(z), \;\;\;
\frac{da_{+}}{dz}   =wq(z)S_{w}(z),
\end{equation}
\[
\frac{da_{-}}{dz}   =\int\limits_{-\infty}^{\infty}\frac{dx}{(ke^{-x}%
+\bar{k}e^{x})^{2}}\left[\frac{i\alpha}{2}(ke^{x}+\bar{k}e^{-x}) +4wx(k^{3}%
e^{x}+\bar{k}^{3}e^{-x})\right][s(x)e^{-i\psi}-\bar{s}(-x)e^{i\psi}].
\noindent
\]
Since the soliton velocity and width enter eq.~(\ref{10}) in the combination $vw^{-1}%
$, we find from~(\ref{9}) the noise source $S_{vw^{-1}}$:%
\begin{eqnarray}
\frac{d}{dz}(\frac{v}{w})&=&-2k^{3}(v+\frac{i}{w})\int\limits_{-\infty}^{\infty
}dx\frac{e^{x}}{(ke^{-x}+\bar{k}e^{x})^{2}}[s(x)e^{-i\psi}+\bar{s}
(-x)e^{i\psi}] \nonumber \\
&-&2\bar{k}^{3}(v-\frac{i}{w})\int\limits_{-\infty}^{\infty}dx\frac{e^{-x}%
}{(ke^{-x}+\bar{k}e^{x})^{2}}[s(x)e^{-i\psi}+\bar{s}(-x)e^{i\psi}]\equiv
S_{vw^{-1}}(z) . \label{12}
\end{eqnarray}
Finally, note that the mean-square fluctuation of the quantity $B$ caused by
the noise source $S_{B}$ is determined by
\[
\langle\bar{B}(z)B(z^{\prime})\rangle=\langle\int\limits_{0}^{z}d\zeta\bar
{S}_{B}(\zeta)\int\limits_{0}^{z^{\prime}}d\zeta^{\prime}S_{B}(\zeta^{\prime
})\rangle .
\]
Now we have everything to calculate the mean-square fluctuation of
the soliton displacement.

\section{Results}
It is seen from~(\ref{10}) that the mean-square fluctuation of the
soliton  displacement is produced by both the noise source
$S_{a}$ driving the displacement directly and the fluctuations of
the combination $vw^{-1}.$ We consider noise sources as
independent, so the fluctuations they produce are
additive. This gives%
\[
\langle q(L)q(L)\rangle=\langle q(L)q(L)\rangle_{a}+\langle
q(L)q(L)\rangle
_{vw^{-1}}.%
\]
Here%
\begin{equation}
\langle q(L)q(L)\rangle_{vw^{-1}}=\int\limits_{0}^{L}dz%
\int\limits_{0}^{L}dz^{\prime}\langle\frac{v}{w}(z)\frac{v}%
{w}(z^{\prime})\rangle . \label{13}
\end{equation}
and%
\[
\langle q(L)q(L)\rangle_{a}=\int\limits_{0}^{L}dz\int\limits_{0}%
^{L}dz^{\prime}\langle\bar{S}_{a}(z)S_{a}(z^{\prime})\rangle .
\]
In turn, the correlation function for the velocity-to-width ratio is
determined by the noise source $S_{vw^{-1}}$(12):%
\begin{equation}
\langle\frac{v}{w}(z)\frac{v}{w}(z^{\prime})\rangle
=\int\limits_{0}^{z}d\zeta\int\limits_{0}^{z^{\prime}}%
d\zeta^{\prime}\langle\bar{S}_{vw^{-1}}(\zeta)S_{vw^{-1}}(\zeta^{\prime
})\rangle . \label{14}
\end{equation}
Straightforward calculation by means of eqs.~(\ref{3}) and~(\ref{12}) gives%
\begin{equation}
\langle\bar{S}_{vw^{-1}}(z)S_{vw^{-1}}(z^{\prime})\rangle=AF(\alpha,v,\gamma
)\delta(z-z^{\prime}) \label{15}
\end{equation}
with%
\begin{eqnarray}
F(\alpha,v,\gamma)&=&\frac{4}{\alpha^{2}}(1-\alpha
v)\{(1-2\alpha v)[(1-\alpha v)\sec^{2} 2\gamma-2(1-2\alpha v)] \nonumber \\
&+& ([(1-\alpha v)\sec^{2}2\gamma-2(1-2\alpha v)]^{2}+(1-2\alpha
v)^{2}\sec ^{2}2\gamma)\gamma\cot2\gamma\} . \label{16}
\end{eqnarray}
In the course of calculation, we ignore, because of smallness of
noise sources, small variations of the parameter $k$. Thereby,
the parameters $V,w$ and $\gamma$ refer to their initial values.
(Remind that distributed gain compensates exactly for fiber
losses). Besides, we express the soliton width $w$ in terms of $v$
and $\gamma$ due to the relation
\begin{equation}
w(1-\alpha v)=\alpha\cot2\gamma . \label{17}
\end{equation}
Finally, we obtain from~(\ref{14}) and~(\ref{13}) the correlation
function for the soliton displacement driven by fluctuations
of the velocity-to-width ratio:%
\begin{equation}
\langle
q(L)q(L)\rangle_{vw^{-1}}=\frac{1}{3}AF(\alpha,v,\gamma)L^3.
\label{18}
\end{equation}
Similar calculation for the $S_{a}$-driven correlation $\langle
q(L)q(L)\rangle_{a}$ in accordance with~(\ref{11}) displays the
linear growth with $L$. Hence, for large propagation distances the
correlation function~(\ref{18}) provides the main contribution to
the total mean-square fluctuation of the soliton displacement.

\begin{figure}
\begin{center}
\epsfig{file=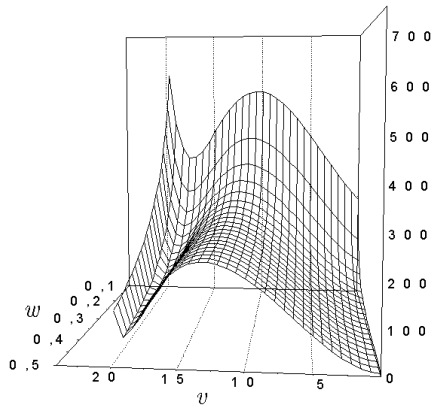,height=6cm,width=8cm}
\caption{Profile of the
function $F(\alpha,v,\gamma)$ determining the correlation
function for the soliton velocity-to-width ratio for
$\alpha=0.05$. The soliton width $w$ is related with the parameter
$\gamma$ through ~(\ref{17}).} \label{f.1}
\end{center}
\end{figure}
\begin{figure}
\centering \mbox{ \epsfig{file=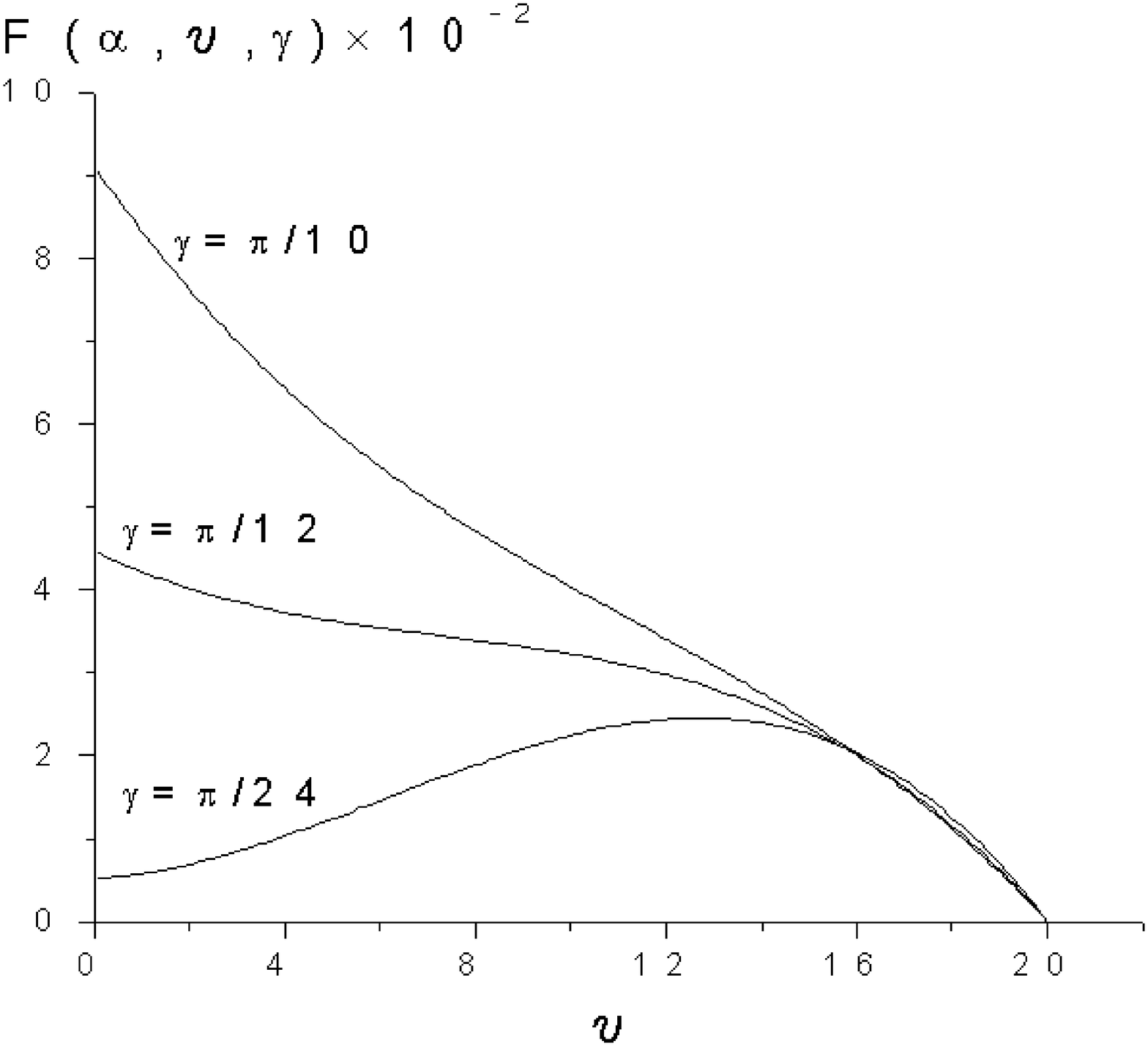,height=6cm,width=6cm}}
\epsfig{file=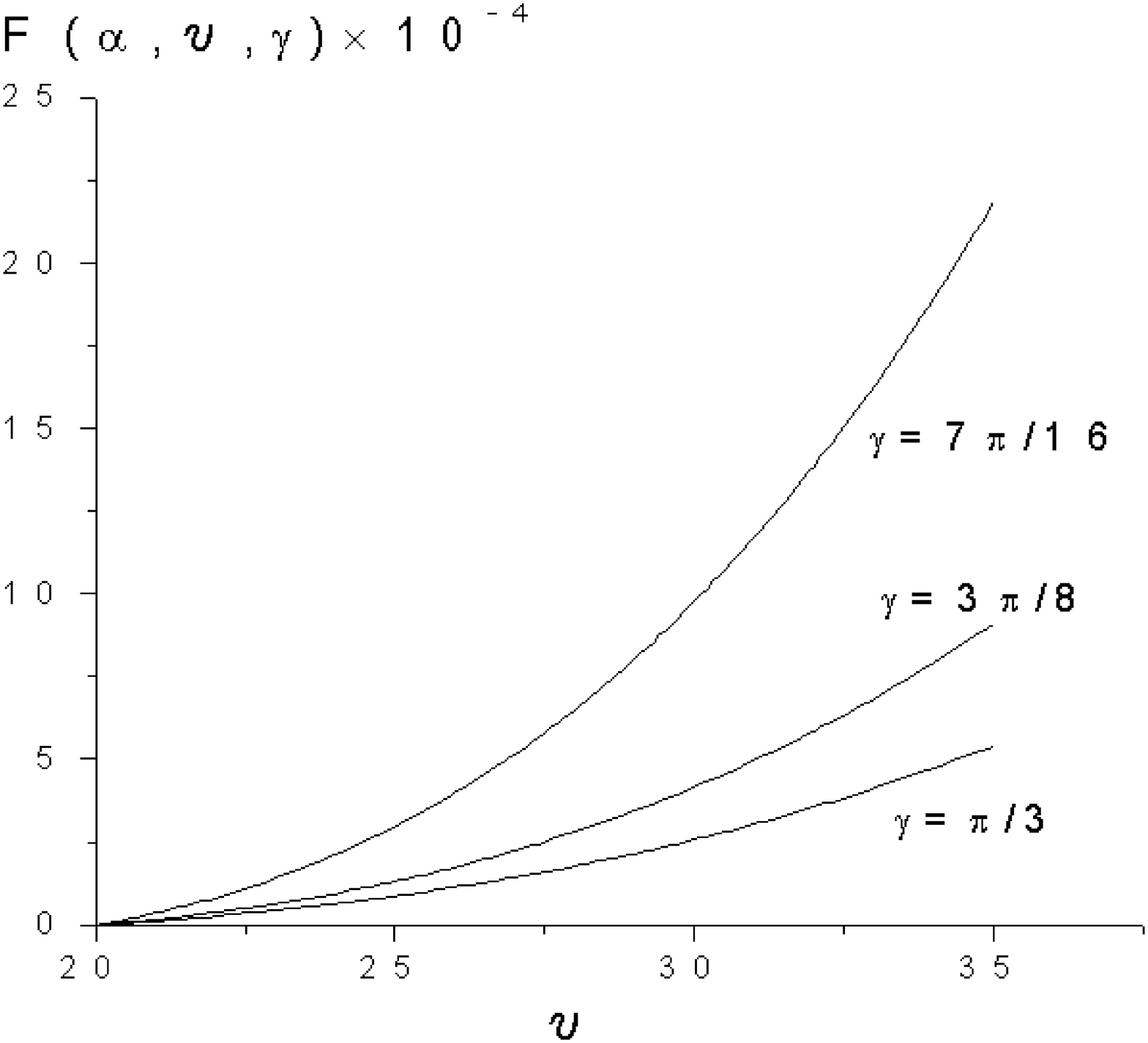,height=6cm,width=6cm} \caption{Typical
behavior of the function $F(\alpha,v,\gamma)$ versus $v$ for
different $\gamma$ and $\alpha=0.05$; (a) $v<\alpha^{-1}$, (b)
$v>\alpha^{-1}$.} \label{f.2}
\end{figure}

\section{Discussion}

The above dependences on $L$ of the mean-square MNLS soliton
fluctuations coincide completely with those in the case of the
NLS soliton~\cite{1,9}. Nevertheless, there is a substantial
difference between NLS and MNLS equations. While the correlation
functions for the NLS soliton do not contain soliton parameters
(in dimensionless units), these parameters enter explicitly into
the r.h.s. of~(\ref{15}) for the MNLS equation. Therefore, we can
reduce the Gordon-Haus jitter for the MNLS soliton by varying
soliton parameters. In Fig. 1 we demonstrate this effect in terms
of variables $v$ and $w$. It is seen that the function~(\ref{16})
smoothly decreases when $v$ approaches $\alpha^{-1}$.

The role of the optical energy $\gamma$ is depicted on Fig. 2. As
follows from~(\ref{17}), we have $0<\gamma<\pi/4$ for
$v<\alpha^{-1}$ and $\pi/4<\gamma<\pi/2$ for $v>\alpha^{-1}$. It
is seen that the function~(\ref{16}) grows significantly for all
$\gamma$ for $v>\alpha^{-1}$, while the same growth for
$v<\alpha^{-1}$\ is displayed to a lesser extent, including
non-monotone behavior for small $\gamma$.

For slightly different definition of the soliton position as
$x=-\frac{1}{w}(t-q(z))$ the function $F(\alpha,v,\gamma)$ takes
more simple form,
\[
F(\alpha,v,\gamma)=2(1-\alpha v)\{1-\cot^22\gamma+\gamma[(1-\tan^2
2\gamma)^2+{\rm sec}^22\gamma]\cot^32\gamma\},
\]
but with the same qualitative behavior.

As regards the applicability of the above result on the soliton
jitter suppression to actual femtosecond optical solitons, we
believe that the MNLS equation is the true integrable model to
start with the account for the third-order dispersion and the
Raman self-frequency shift. It should be noted that in the
adiabatic approximation the third-order dispersion does not
contribute to the MNLS soliton velocity and width, just as for
the NLS soliton~\cite{16}. The Raman effect contribution can be
calculated additively as an extra perturbation to the MNLS
soliton, within the framework of the MNLS soliton perturbation
theory~\cite{13}. In so doing, a value of $\alpha$ should be
carefully reconciled with the specific conditions of femtosecond
pulse propagation. It is worth noting that there exists a
possibility~\cite{17} to partially compensate for the Raman
self-frequency shift effect, as applied to femtosecond optical
pulses. Corresponding results will be published elsewhere.

In conclusion, we have shown that for solitons of the MNLS
equation which can serve as the integrable model for the
description of ultrashort soliton dynamics, the Gordon-Haus
jitter can be significantly reduced by means of matching soliton
parameters alone. We stress that this reduction is
non-perturbative with respect to $\alpha$, i.e., it cannot be
revealed in the framework of the NLS equation with
$\alpha$-dependent term considered as a perturbation. Though the
above result is valid within the adiabatic approximation of the
MNLS soliton perturbation theory, we believe it describes the
main features of this phenomenon for ultrashort optical pulses.


\newpage
\begin{thebibliography}{0}

\bibitem{1}
  J.P. Gordon  and H.A Haus. {\it Opt. Lett.} {\bf11}, 665 (1986).

\bibitem{2}
  A. Mecozzi, J.D. Moores, H.A. Haus and Y. Lai.
  {\it Opt. Lett.} {\bf16}, 1841 (1991).

\bibitem{3}
  Y. Kodama and A. Hasegawa.
  {\it Opt. Lett.} {\bf 17}, 31 (1992).

\bibitem{4}
  W. Forysiak, K.J. Blow and N.J. Doran.
  {\it Electron. Lett.} {\bf29}, 1225 (1993).

\bibitem{5}
  N.J. Smith, W. Forysiak and N.J. Doran.
  {\it Electron. Lett.} {\bf32}, 2085 (1996).

\bibitem{6}
  S. Kumar and F. Lederer.
  {\it Opt. Lett.} {\bf 22}, 1870 (1997).

\bibitem{7}
  C. Pare, A. Villeneuve, P.A. Belanger and N.J. Doran.
  {\it Opt. Lett.} {\bf21}, 459 (1996).

\bibitem{8}
  V.V. Kozlov and A.B. Matsko.
  {\it J. Opt. Soc. Amer. B} {\bf16}, 519 (1999).

\bibitem{9}
  H.A. Haus and W.S. Wong.
  {\it Rev. Mod. Phys.} {\bf68}, 423 (1996).

\bibitem{10}
  G.P. Agrawal.
  {\it Nonlinear Fiber Optics}
  (Academic Press, San Diego, CA), 1995.


\bibitem{11}
  M. Wadati, K. Konno and Y.H. Ichikawa.
  {\it J. Phys. Soc. Jpn.} {\bf46}, 1965 (1979).

\bibitem{12}
  K. Ohkuma, Y.H. Ichikawa and Y. Abe Y.
  {\it Opt. Lett.} {\bf12}, 516 (1987).

\bibitem{13}
  V.S. Shchesnovich and E.V. Doktorov.
  {\it Physica D} {\bf129}, 115 (1999).

\bibitem{14}
  V.S. Gerdjikov and M.I. Ivanov.
  {\it Bulgarian J. Phys.} {\bf10}, 13 (1983).

\bibitem{15}
  J.P. Gordon.
  {\it Opt. Lett.} {\bf8}, 596 (1983).

\bibitem{16}
  J.N. Elgin.
  {\it Phys. Rev. A} {\bf47}, 4331 (1993).

\bibitem{17}
  A.A. Afanas'ev, E.V. Doktorov, R.A. Vlasov and V.M. Volkov.
  {\it Optics Comm.} {\bf153}, 83 (1998).

\end{thebibliography}
\end{document}